\documentclass[preprint,onecolumn,nofootinbib]{revtex4}
\pdfoutput=1
\usepackage[colorlinks=true,linkcolor=blue,urlcolor=blue,filecolor=black,citecolor=red,pdfstartview=FitV,pdftitle={},pdfsubject={},pdfkeywords={},pdfpagemode=None,bookmarksopen=true]{hyperref}
\usepackage{graphicx}
\usepackage{amsmath}
\usepackage{amsfonts}
\usepackage{amssymb,ulem}
\usepackage{color}%
\usepackage{CJK}
\usepackage{dcolumn}
\newcommand{\f}{\begin{equation}}
\newcommand{\ff}{\end{equation}}
\newcommand{\fa}{\begin{eqnarray}}
\newcommand{\ffa}{\end{eqnarray}}

\begin{document}
\title{Shear viscoelasticity in anisotropic holographic axion model}
\author{Lei Li}
\author{Wei-Jia Li}
\thanks{weijiali@dlut.edu.cn (corresponding author)}
\affiliation{
Institute of Theoretical Physics, School of Physics, Dalian
University of Technology, Dalian 116024, China}
\author{Xiao-Mei Kuang}
\thanks{xmeikuang@yzu.edu.cn (corresponding author)}
\affiliation{
Center for Gravitation and Cosmology, College of Physical Science and Technology, Yangzhou University, Yangzhou 225009, China}
\begin{abstract}
In this work, we investigate the shear elasticity and the shear viscosity in a simple holographic axion model with broken translational symmetry and rotational symmetry in space via the perturbation computation. We find that, in the case of spontaneous symmetry breaking, the broken translations and anisotropy both enhance the shear elasticity of the system. While in all cases, the broken symmetries introduce a double suppression on the shear viscosity, which is in contrast to the result from the study of the p-wave holographic superfluid where the shear viscosity is enhanced when the rotational symmetry is broken spontaneously. 

\end{abstract}
\maketitle
\tableofcontents
\section{Introduction}
During the past two decades, the AdS/CFT correspondence has become a powerful tool for studying the real-time dynamics of strongly coupled systems \cite{Ammon_Erdmenger_2015,Zaanen_Liu_Sun_Schalm_2015,Hartnoll:2016apf,Baggioli:2019rr}. It provides us a geometric approach to calculate transport coefficients of boundary systems explicitly in the AdS bulk by considering various black hole solutions which can dissipate the surrounding fluctuations outside their horizons. One of the most important discoveries by this method is that for a wide class of interacting systems, there exists a lower bound on the shear viscosity. And especially, for those having a gravity dual that can be described by Einstein gravity, the ratio of shear viscosity to entropy density $\eta/s$ meets a universal value $\hbar/4\pi k_B$ which is called the Kovtun-Son-Starinet (KSS) bound \cite{Policastro:2001yc,Kovtun:2003wp,Kovtun:2004de}.
Remarkably, such a finding in theory also explains why the ratio of shear viscosity to entropy density of the quark-gluon-plasma (QGP) is much smaller when comparing against the result from the perturbative calculations of quantum field theory \cite{Arnold:2003zc,Huot:2006ys}. This bound has been tested in various experiments of different realistic systems \cite{Luzum:2008cw,Schafer:2009dj,Cremonini:2011iq,Nagle:2011uz,
Shen:2011eg}.

Nevertheless, it was found that $\eta/s$ can be corrected when the finite $N$ effect is considered. In this case, the viscosity bound can be slightly  pushed down
\cite{Brigante:2008gz,Brigante:2007nu}.\footnote{More generally, the violation of KSS bound was investigated in higher order corrections \cite{Iqbal:2008by,Cai:2008ph,Cai:2008in,Brustein:2008cg,Cai:2009zv,Kats:2007mq,Myers:2010jv,Feng:2015oea,Wang:2016vmm,Buchel:2024umq}.} In addition, the viscosity bound can be strongly violated when matter fields are included in. A numerous specific examples have shown that this can happen when the rotational symmetry or the translational symmetry is broken (Examples of these two classes can be seen in \cite{Natsuume:2010ky,Erdmenger:2010xm,Rebhan:2011vd,Mamo:2012sy,Giataganas:2013lga,Jain:2014vka,Critelli:2014kra,Ge:2014aza,Jain:2015txa,Ge:2015owa,Landsteiner:2016stv,Finazzo:2016mhm,samanta:2016pic,Jeong:2017rxg,Giataganas:2017koz} and \cite{Davison:2014lua,Baggioli:2016rdj,Burikham:2016roo,Liu:2016njg,Ling:2016ien,Hartnoll:2016tri,Alberte:2016xja,Ling:2016yxy,Cisterna:2017jmv,Baggioli:2018bfa,Andrade:2019zey,Zhou:2019xzc,Baggioli:2020ljz,Wang:2021jfu,Baggioli:2021tzr,Xia:2024gba}, respectively.). In most of these cases, the way of breaking the symmetries is explicit. However, a recent holographic study on the p-wave superfluid model shows that the ratio $\eta/s$ can be enhanced even in an anisotropic system if the rotations are broken in a spontaneous manner \cite{Baggioli:2023yvc}. So far, this model appears to be the only known anisotropic case that obeys the KSS bound.

In this work, we investigate the shear elasticity and the shear viscosity of an anisotropic holographic axion model at the perturbative level. This model allows us to realize the broken translations and the broken rotations explicitly and/or spontaneously via different setups. The boundary dual of this model is supposed to be certain viscoelastic strongly-coupled solids which exhibit both elastic responses as normal crystals and viscous damping as fluids. We expect that such a holographic tool will deepen our understanding of viscoelastic properties of these complex materials in real world.
In particular, we would like to investigate the fate of the KSS bound in the case of spontaneous breaking of the symmetries. Our result shows that, in contrast to the p-wave superfluid, the KSS bound is always violated in our model.

\section{Anisotropic black hole solutions}
We consider a simple holographic axion model \cite{Baggioli:2014roa,Alberte:2016xja,Alberte:2015isw,Baggioli:2016rdj,Andrade:2019zey,Baggioli:2021xuv,Baggioli:2023dfj} in 4-dimensional spacetime, of which the action is given by
\begin{equation}
S=\int \mathrm{d}^{4}x\,\sqrt{-g} \left(R+\frac{6}{L^2}-\lambda^2 \, V(X)\right),\quad X\equiv\frac{1}{2}\partial_\mu \phi^i\partial^\mu \phi^i,\,\,i=x,y
\end{equation}
where we have adopted the convention that $16\pi G\equiv 1$, $R$ is the Ricci scalar, $\lambda^2$ is the effective coupling constant  with the dimension of mass square \footnote{The physical coupling  in front of $V(X)$, $\frac{\lambda^2}{16\pi G}$, should be dimensionless, which requires the dimension $[\lambda^2]=[G]=[M^2]$.} , $V$ is a general function of $X$, and $L$ is the AdS radius. To break the translations of the boundary system, the profiles of the scalars in the bulk should be chosen as  
\begin{equation}\label{profile}
\phi^i=\mathcal{M}^{i}_{j}\,x^j,\quad \mathcal{M}^{i}_{j}=\begin{pmatrix}
    k_x&  0 \\
   0             &   k_y
   \end{pmatrix}.
\end{equation}
Then, the general AdS black hole ansatz should be taken as
\begin{eqnarray}
\mathrm{d}s^2&&=g_{tt}\mathrm{d}t^2+g_{rr}\mathrm{d}r^2+g_{xx}\mathrm{d}x^2+g_{yy}\mathrm{d}y^2\\ \nonumber
&&=-A(r)\mathrm{d}t^2+\frac{\mathrm{d}r^2}{A(r)}+B(r)\mathrm{d}x^2+C(r)\mathrm{d}y^2,
\end{eqnarray}
of which the metric components, at the AdS boundary $r\rightarrow \infty$, should satisfy
\begin{eqnarray}
&&A(r\rightarrow \infty)=B(r\rightarrow \infty)=C(r\rightarrow \infty)=\frac{r^2}{L^2}.
\end{eqnarray}
The location of the horizon $r_h$ is then defined by $A(r=r_h)=0$. For simplicity, we will set $L$ to be one \footnote{This means that all the dimensional quantities are rescaled by $L$. For instance, $\lambda\rightarrow \lambda L$ which becomes a dimensionless quantity.}, and consider a class of specific models with  
\begin{equation}
V(X)=X^n.
\end{equation}
By considering $\phi^i=\bar{\phi^i}+\delta \phi^i$ and expanding the action to the second order in $\delta \phi^i$, we obtain
\begin{equation}
\frac{1}{2}V'(\bar{X})\partial_{\mu}\delta\phi^i\partial^{\mu}\delta\phi^i+\bar{X}V''(\bar{X})(\partial_{i}\delta \phi^i)^2+\cdots.
\end{equation} 
The absence of ghost requires monotonic potentials
\begin{equation}\label{no ghost}
V'(\bar{X})>0,
\end{equation} 
which gives us $n>0$ \cite{Baggioli:2014roa}. 
For general $n>0$, the axions behave like
\begin{equation}\label{UVEXP}
\phi^i(r,x^\mu)=\phi_{(0)}^i(x^\mu)+\phi_{(1)}^i(x^\mu)\,r^{2n-5}+\cdots
\end{equation}
near the AdS boundary, where the $r$-independent term corresponds to the profiles of the axions (\ref{profile}).  Following the standard quantization, the leading mode of the axion fields sets the external source for the dual scalar operators on boundary. And the expectation values of the scalar operators correspond to the subleading mode of the axion fields. For $n<5/2$, the $r$-independent term in  (\ref{UVEXP}) dominates over the expansion and plays role of the external source. Then,  the profile (\ref{profile}) with $k_x\neq k_y$ implies that the translations and the rotations are both broken explicitly. While, for $n>5/2$, the $r$-independent term becomes subleading and the breaking of symmetries is spontaneous \cite{Alberte:2017oqx}. In the next section, we will investigate how the broken translations and the broken rotations affect the shear viscosity-entropy density ratio in these two scenarios.

In the following, we solve the background solutions perturbatively in two cases:
\subsection*{One-axion case }
First, let us consider the single axion case where $k_x=0,\,k_y\neq0$ are imposed without loss of generality. Note that the low anisotropic regime can be achieved by setting $\lambda \ll 1$ and fixing the value of $k_y$. Then, it is found that the background can be expressed perturbatively up to the leading order for $\mathcal{O}(\lambda^2)$,
\begin{eqnarray}\label{generalnbg}
&&A(r)=r^2\left(1-\frac{r_h^3}{r^3}\right)+\lambda^2\, a_1^{(n)}(r)+\mathcal{O}(\lambda^4),\\
&&B(r)=r^2+\lambda^2\, b_1^{(n)}(r)+\mathcal{O}(\lambda^4),\\
&&C(r)=r^2-\lambda^2\, b_1^{(n)}(r)+\mathcal{O}(\lambda^4),
\end{eqnarray}
with
\begin{eqnarray}\label{a1}
&&a_1^{(n)}(r)=\frac{k_y^{2n}}{2^{n+1}\,(2n-3)}\frac{r^{3-2n}-r_h^{3-2n}}{r},\\ \label{b1}
&&b_1^{(n)}(r)=\frac{n\,k_y^{2n}\,r^{2}\,r_h^{-2n}}{3 \cdot 2^{n}\,(2n-3)}\left[\text{log}\left(1-\frac{r_h^3}{r^3}\right)+\text{Re}\,\mathcal{B}_0\left(\frac{r^3}{r_h^3};1-\frac{2n}{3},0\right)\right]+\gamma _{1(n)} \,r^2,
\end{eqnarray}
where \text{Re}$\mathcal{B}_0\left(x;1-\frac{2n}{3},0\right)$ represents the regular sector of the real part of  the incomplete beta function $\mathcal{B}\left(x;1-\frac{2n}{3},0\right)$ and the constant $\gamma _{1(n)}$ is 
\begin{eqnarray}\label{gamma}
&&\gamma _{1(n)} \equiv
\begin{cases}
\frac{n\,\pi\, k_y^{2 n}\,r_h^{-2 n}}{3\, (2 n-3) \,2^n}\,\text{cot}\left(\frac{2 \pi  n}{3}\right), & n\neq\frac{3}{2}\textbf{z},\\
0, & n= \frac{3}{2}(\textbf{z}+1),
\end{cases}
\end{eqnarray}
with $\textbf{z}$  denoting positive integers. Here, the special case $n =\frac{3}{2}$ has been excluded. It should be mentioned that the above results (\ref{a1})-(\ref{gamma}) are obtained by choosing the integral constants properly, which ensures the regularity of the background solution in the bulk and does not disrupt the asymptotic AdS geometry. For more details about the derivation, one refers to Appendix \ref{Appendix A}. Then, the Hawking temperature can be read off as
\begin{eqnarray}\label{one axion temperature}
T=\frac{A'(r_h)}{4\pi}=\frac{3\,r_h}{4\pi}-\lambda^2\,\frac{{k_y}^{2n}{r_h}^{1-2n}}{2^{n+3}\,\,\pi}+\mathcal{O}(\lambda^4).
\end{eqnarray}

While for $n=\frac{3}{2}$, one should solve it separately and obtain the corrections in the background metric as follows 
\begin{eqnarray}\label{}
&&a_1^{(n)}(r)=\frac{k_y^3 }{4 \sqrt{2}\,r}\text{log}\left(\frac{r_h}{r}\right),\\
&&b_1^{(n)}(r)=-\frac{r^2\,k_y^3}{72 \sqrt{2} r_h^3}\left[\pi ^2+27 \left(\text{log} \frac{r}{r_h}\right)^2+6\, \text{Li}_2\left(1-\frac{r^3}{r_h^3}\right)\right],
\end{eqnarray}
with the polylogarithm function $\text{Li}_{2}(x)$. And the Hawking temperature becomes 
\begin{eqnarray}
&&T=\frac{3 \,r_h}{4 \pi }-\lambda^2 \frac{ k_y^3}{16\, \sqrt{2}\, \pi \, r_h^2}+\mathcal{O}(\lambda^4).
\end{eqnarray}

\subsection*{Two-axion case}

Now, we consider that there are two axions along both $x$ and $y$ directions, which means to  set $k_x,\,k_y\neq0$. For the general cases, one can assume that $k_x\neq k_y$. In the low anisotropic regime, we still have
\begin{eqnarray}
&&A(r)=r^2\left(1-\frac{r_h^3}{r^3}\right)+\lambda^2\, a_2^{(n)}(r)+\mathcal{O}(\lambda^4),\\
&&B(r)=r^2+\lambda^2\,b_2^{(n)}(r)+\mathcal{O}(\lambda^4),\\
&&C(r)=r^2-\lambda^2\,b_2^{(n)}(r)+\mathcal{O}(\lambda^4),
\end{eqnarray}
where 

\begin{align}
a_2^{(n)}(r)=&\frac{{(k_x^2+k_y^2)}^{n}}{2^{n+1}(2n-3)}\frac{r^{3-2n}-r_h^{3-2n}}{r},\\
b_2^{(n)}(r)=&\frac{n\,(k_x^2+k_y^2)^{n-1}\,r^{2}\,r_h^{-2n}}{3\cdot2^{n}\,(2n-3)}(k_y^2-k_x^2)\left[\text{log}\left(1-\frac{r_h^3}{r^3}\right)+\text{Re}\,\mathcal{B}_0\left(\frac{r^3}{r_h^3};1-\frac{2n}{3},0\right)\right]
\\ \nonumber
&+\gamma _{2(n)} \,r^2,
\end{align}
with
\begin{eqnarray}
\gamma _{2(n)} \equiv
\begin{cases}
\frac{n\,\pi\, (k_y^2-k_x^2)\,(k_x^2+k_y^2)^{n -1}\,r_h^{-2 n}}{3\, (2 n-3) \,2^n}\,\text{cot}\left(\frac{2 \pi  n}{3}\right), & n\neq\frac{3}{2}\textbf{z},\\
0, & n=\frac{3}{2}(\textbf{z}+1),
\end{cases}
\end{eqnarray}
for $n\neq\frac{3}{2}$. The Hawking temperature then is given by
\begin{eqnarray}\label{two axion temperature}
T=\frac{A'(r_h)}{4\pi}=\frac{3\,r_h}{4\pi}-\lambda^2\,\frac{({k_x^2+k_y^2)}^{n}{r_h}^{1-2n}}{2^{n+3}\,\,\pi}+\mathcal{O}(\lambda^4).
\end{eqnarray}

For $n=\frac{3}{2}$, the results turn into
\begin{eqnarray}
&&a_2^{(n)}(r)=\frac{(k_x^2+k_y^2 )^{3/2}}{4 \sqrt{2}\,r}\text{log} \left(\frac{r_h}{r}\right),\\
&&b_2^{(n)}(r)=-\frac{r^2\,(k_y^2-k_x^2)\,\sqrt{k_x^2+k_y^2}}{72 \sqrt{2} r_h^3}\left[\pi ^2+27 \left(\text{log} \frac{r}{r_h}\right)^2+6\, \text{Li}_2\left(1-\frac{r^3}{r_h^3}\right)\right],\\\nonumber 
\end{eqnarray}
and
\begin{eqnarray}
&&T=\frac{3\,r_h}{4 \pi }-\lambda^2\, \frac{ \left(k_x^2+k_y^2\right){}^{3/2}}{16\, \sqrt{2}\,\pi \,r_h^2}+\mathcal{O}(\lambda^4).
\end{eqnarray}

In addition, one can easily check that, for the isotropic case (i.e., $k_x=k_y$), $b_2^{(n)}(r)$ is vanishing, which is consistent with the known results from previous studies on holographic solids \cite{Alberte:2016xja,Andrade:2019zey,Alberte:2015isw,Baggioli:2019abx,Baggioli:2023dfj}.

\section{Viscoelastic response}
In the hydrodynamic limit (low frequency and small momentum), the retarded Green function of the stress tensor at zero momentum can be expanded as \cite{Ammon:2019wci}
\begin{eqnarray} 
G^{R}_{T_{xy}T_{xy}}(\omega,\boldsymbol{p}=0)=\mu_{xy}-\mathrm{i}\,\omega\,\eta_{xy}+\mathcal{O}(\omega^2),
\end{eqnarray}
where the non-dissipative part $\mu_{xy}=\lim\limits_{\omega\rightarrow 0}\text{Re}\,G^R_{T_{xy}T_{xy}}(\omega,\boldsymbol{p}=0)$ is interpreted as the shear elastic modulus, and the dissipative coefficient $\eta_{xy}$ in the imaginary part  is associated to entropy production and should be understood as the shear viscosity.

In the holographic framework, to compute $G^{R}_{T_{xy}T_{xy}}(\omega,\boldsymbol{p}=0)$, we need to introduce time-dependent perturbations $\delta g_{xy}$ upon the background in the bulk. Note that this metric fluctuation is decoupled from other possible fluctuations for zero $\boldsymbol{p}$ because $\delta g_{xy}$ is the solo fluctuation with tensor channel in this sector. For $g_{ij}=\bar{g}_{ij}+\delta g_{ij}$ (From now on, we will use the bar to denote background metric we obtained in last section.), taking the Fourier transformation $\delta g^i_j(t,r, x^i)\sim h^i_j(r)\,\mathrm{e}^{-\mathrm{i}\omega t}$, we achieve the equations of $h^x_y$  and $h^y_x$ as follows\footnote{We here do not adopt the Einstein convention for the spatial indeces in the linearized equations.}:
\begin{eqnarray}
\frac{1}{\sqrt{-g}}\partial_r\left(\sqrt{-g}Z^j_{i}(r)\bar{g}^{rr}\partial_r h^{i}_{j}\right)-\omega^2\bar{g}^{tt}Z^j_{i}(r) h^{i}_{j}= {m^i_g}(r)^2 Z^j_{i}(r)h^{i}_{j}, \,\,\,\,\,\,(i=x,y\,\,\,\,j=x,y).
\end{eqnarray}
where $Z^j_i(r)\equiv\bar{g}^{jj}\bar{g}_{ii}$ and the square of the graviton masses ${m^i_g}^2$ match the formula proposed in \cite{Hartnoll:2016tri} which is given by
\begin{eqnarray}
{m^i_g}^2\equiv g^{ii}T_{ii}-\frac{\delta T_{ij}}{\delta {g_{ij}}},
\end{eqnarray}
where the bulk stress tensor in our model is given by
\begin{eqnarray}
T_{MN}=-g_{MN}\lambda^2\,V+\lambda^2\,V'(X)\,\partial_M\phi^i\partial_N\phi^i,
\end{eqnarray}
with $V'\equiv\frac{d V}{dX}$. Obviously, the graviton masses originate from the non-trivial profiles of the scalars in the bulk. For the isotropic case, i.e., $k_x=k_y=k$, we get $Z^j_i=1$ and a unique graviton mass $m_g$. Then, the graviton behaves like a massive scalar in the bulk.

In our anisotropic model, there exist two distinct graviton masses which are
\begin{eqnarray}
{m^i_g}^2=\lambda^2\,k_i^2\,V'(\bar{X})\,\bar{g}^{ii}.
\end{eqnarray}
Therefore, the linearized perturbative equations of $\delta g_{xy}$ can be expressed more explicitly as
\begin{eqnarray}
\partial_r\left(\sqrt{-g}Z^j_{i}\bar{g}^{rr}\partial_r h^{i}_{j}\right)-\omega^2\sqrt{-g}\bar{g}^{tt}Z^j_{i}h^{i}_{j}=\lambda^2 \, k_i^2\sqrt{-g} \bar{g}^{jj}V'h^{i}_{j}.
\end{eqnarray}
Next, we calculate the shear viscoelasticity both in the single axion case and in the two-axion case.
\subsection*{One-axion case}
For $k_x=0$, $k_y\neq0$, the metric perturbation $h^x_y$ satisfies the following equation
\begin{eqnarray}
\partial_r\left(\sqrt{-g}Z^{y}_x\bar{g}^{rr}\partial_r h^{x}_{y}\right)-\omega^2\sqrt{-g}Z^{y}_x\bar{g}^{tt} h^{x}_{y}=0.
\end{eqnarray}
Since the field is massless, one can apply the membrane paradigm which allows us to express the shear viscosity in terms of the horizon data \cite{Kovtun:2003wp,Iqbal:2008by}\footnote{One can also obtain the shear viscosity by solving the equation of $h^y_x$ whose mass makes the computation a little more difficult. However, as is shown in Appendix \ref{Appendix C} that the final result does not depend on which equation we solve.}. For the boundary system with two spatial dimensions, there can only be one shear viscosity $\eta_{xy}$ and one shear modulus $\mu_{xy}$. We will from now on omit the spatial indexes for simplicity. As the result, the shear viscosity to
entropy density ratio can be expressed as
\begin{eqnarray}
\frac{\eta}{s}=\frac{Z^y_x(r_h)}{4\pi}=\frac{1}{4\pi}\frac{\bar{g}_{xx}(r_h)}{\bar{g}_{yy}(r_h)}=\frac{1}{4\pi}\frac{B(r_h)}{C(r_h)}.
\end{eqnarray}
For the small $\lambda$, we obtain that
\begin{eqnarray}\label{one axion eta}
\frac{\eta}{s}=\frac{1}{4\pi}+\lambda^2\,\frac{b_1(r_h)}{2\,\pi\,{r_h}^2}+\mathcal{O}(\lambda^4).
\end{eqnarray}
In Appendix \ref{Appendix A}, we have shown that $b_1(r_h)<0$, which implies that, whether spontaneous or explicit, the rotational symmetry breaking always results in a violation of the KSS bound. In addition, the shear modulus in this case is zero due to the vanishing graviton mass.

\subsection*{Two-axion case}
For $k_x,\, k_y\neq0$, the rotational symmetry is entirely broken. In this case, $h^x_{y}$ becomes massive which satisfies the following equation
\begin{eqnarray}
\partial_r\left(\sqrt{-g}Z^{y}_x \bar{g}^{rr}\partial_r h^{x}_{y}\right)-\omega^2\sqrt{-g}Z^{y}_x \bar{g}^{tt} h^{x}_{y}=\lambda^2\,k_x^2\sqrt{-g} \bar{g}^{yy}V'h^{x}_{y}.
\end{eqnarray}
One can show that the graviton mass introduces a non-zero real part of $G^{R}_{T_{xy}T_{xy}}(\omega,\boldsymbol{p}=0)$. Unlike the previous case, both $\mu$ and $\eta$ now depend on the full geometry. The membrane paradigm is no longer applicable. However, for small $\lambda$, one can analytically calculate them order by order. The details have been shown in Appendix \ref{Appendix B}. As the result, $\mu$ and $\eta/s$ are
\begin{eqnarray}\label{two-axion mu1}
\mu=\lambda^2\frac{n\,k_x^2\,\left({k_x^2+k_y^2}\right)^{n-1}}{2^{n-1}\, (2n-3)}r_h^{3-2n}+\mathcal{O}(\lambda^{4}), \quad n>\frac{3}{2}
\end{eqnarray}
and
\begin{equation}\label{two-axion eta1}
\frac{\eta}{s}=\frac{1}{4\pi}\frac{B(r_{h})}{C(r_{h})}-
\begin{cases}
\lambda^2\frac{n\,  k_x^2\,\left({k_x^2+k_y^2}\right)^{n-1}}{3\cdot2^{n}\,(2n-3)\, \pi\,r_h^{2n}}\mathcal{H}(\frac{2n-3}{3})+\mathcal{O}(\lambda^{4}),&  n\neq\frac{3}{2}, \\
\lambda^2 \frac{\pi\,k_x^2 \sqrt{k_x^2+k_y^2}}{72\,\sqrt{2}\,r_h^3}+\mathcal{O}(\lambda^{4}),& n= \frac{3}{2},\\
\end{cases}
\end{equation}
where $\mathcal{H}(x)$ is harmonic number.

\begin{figure}[htbp]
    \centering    \includegraphics[width=0.49\linewidth,height=4.7cm]{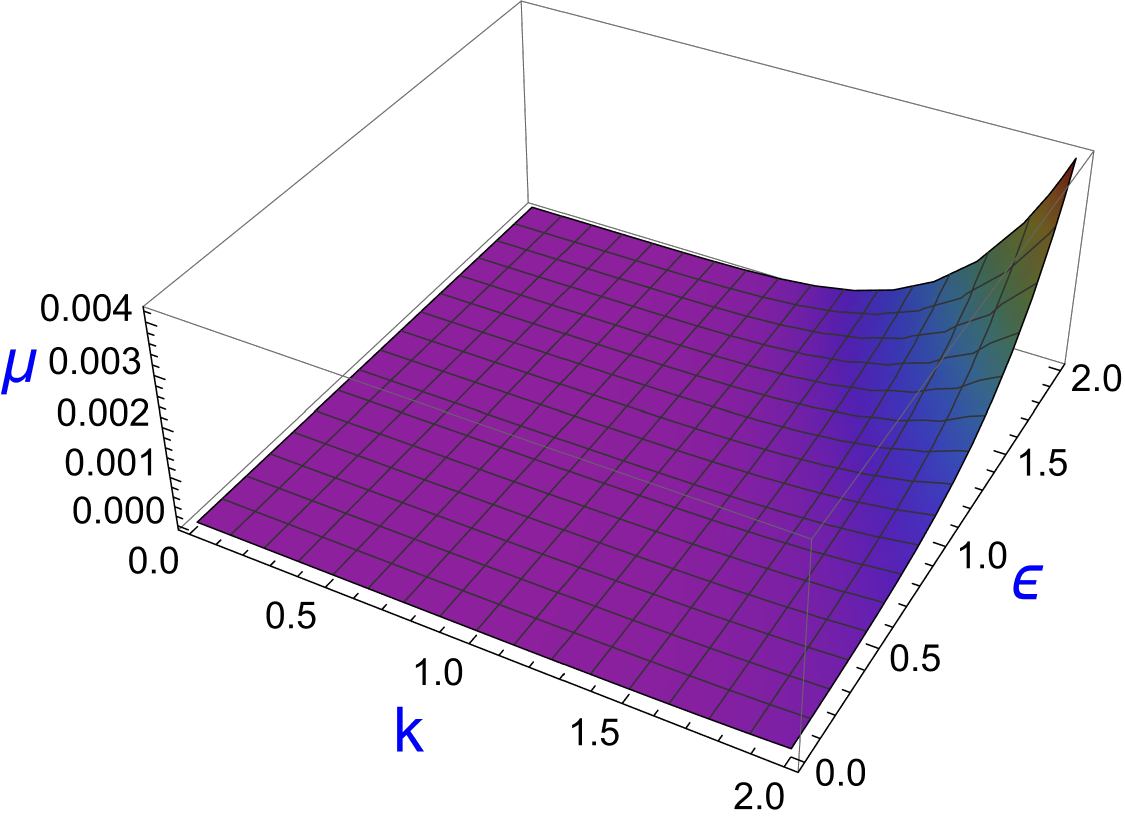} \hspace{0.1cm}  \includegraphics[width=0.49\linewidth,height=4.7cm]{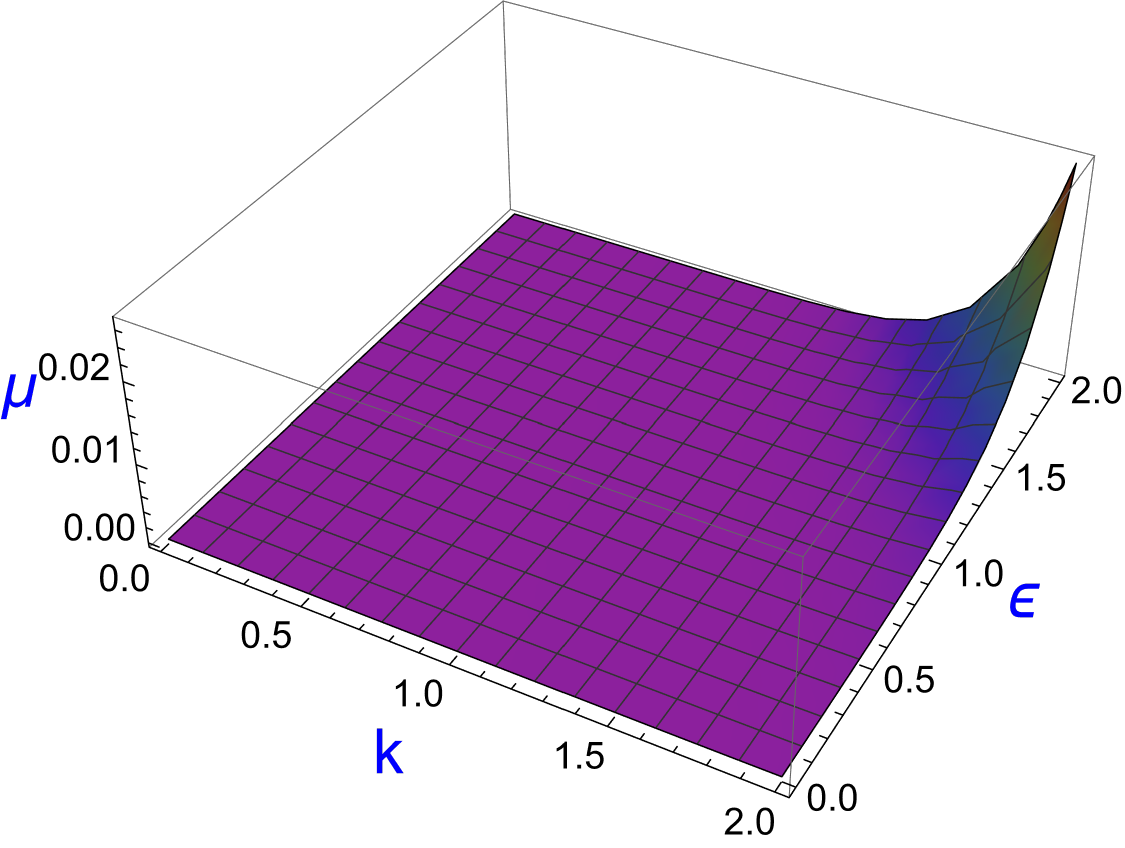}
    \caption{The shear modulus $\mu$ as the function of the $k$ and $\epsilon$, depicted by (\ref{two-axion mu11}). Here, we have fixed $\lambda^2\,L^2=10^{-5}$ and $r_h/L=2$. $\textbf{Left}$: $n=3$. $\textbf{Right}$: $n=5$.  In both cases, the symmetries are spontaneously broken.}
    \label{fig1:MU}
\end{figure}

In order to separate the effects of broken translations and anisotropy in a manifest way, one can rotate the basis spatial coordinates as in \cite{Alberte:2018doe},
\begin{eqnarray}
\phi ^i=k  \left(
\begin{array}{cc}
 \sqrt{\frac{\epsilon ^2}{4}+1} & \frac{\epsilon }{2} \\
 \frac{\epsilon }{2} & \sqrt{\frac{\epsilon ^2}{4}+1} \\
\end{array}
\right)\left(
\begin{array}{c}
 \tilde{x} \\
 \tilde{y} \\
\end{array}
\right),
\end{eqnarray}
where $k$ and $\epsilon$ are two dimensionless parameters characterizing the strengths of translational symmetry breaking and anisotropy, respectively. If we assume that $k_x>k_y$, then
they are related to $k$ and $\epsilon$ as follows,
\begin{eqnarray}
k_x=k\,\left(\sqrt{\frac{\epsilon ^2}{4}+1}+\frac{\epsilon}{2} \right),\\ 
k_y=k\, \left( \sqrt{\frac{\epsilon ^2}{4}+1}-\frac{\epsilon}{2} \right).
\end{eqnarray}
Finally, (\ref{two-axion mu1}) and (\ref{two-axion eta1}) can be re-expressed as
\begin{eqnarray}\label{two-axion mu11}
&&\mu =\lambda^2\frac{n\,k ^{2 n}\,\left(2+\epsilon ^2\right)^{n-1}\,\left(\sqrt{4+\epsilon ^2}+\epsilon \right)^2 }{2^{n+1}\,(2 n-3)\,r_h^{2 n-3}}+\mathcal{O}(\lambda^4) ,\,\,n>\frac{3}{2},
\end{eqnarray}
and
\begin{equation}\label{two-axion eta11}
\frac{\eta}{s}=\frac{1}{4 \pi }-
\begin{cases}
\lambda^2 \,\frac{n\,k ^{2 n} \left(\epsilon ^2+2\right)^n\,\mathcal{H}(\frac{2n-3}{3})}{3\cdot2^{n+1}\,(2 n-3)\,\pi\,r_h^{2 n}}+\mathcal{O}(\lambda^4),&  n\neq\frac{3}{2}, \\
\lambda ^2\,\frac{ \pi\,\left[k ^2 \left(\epsilon ^2+2\right)\right]^{3/2}}{144\,\sqrt{2}\,r_h^3}+\mathcal{O}(\lambda^4),&  n=\frac{3}{2}.\\
\end{cases}
\end{equation}
From the above results, it is obvious to see that $\mu$ just increases monotonically with $k$ and $\epsilon$, and its value is enhanced faster for larger values of $n$. In Appendix \ref{Appendix A}, we show that $\mathcal{H}(\frac{2n-3}{3})<0$ for $0<n< \frac{3}{2}$ and $\mathcal{H}(\frac{2n-3}{3})>0$ for $n>\frac{3}{2}$. Therefore, $\eta/s$ decreases monotonically with $k$ and $\epsilon$ for all the cases. Furthermore, its value is suppressed more significantly for larger values of $n$, which implies that the KSS bound is always violated in the axion model, regardless of whether the symmetries are broken explicitly or spontaneously. In FIG.\ref{fig1:MU}-FIG.\ref{fig3:eta}, we plot $\mu$ and $\eta/s$ as functions of $k$ and $\epsilon$ for different cases.

\begin{figure}[htbp]
    \centering
    \includegraphics[width=0.49\linewidth,height=4.7cm]{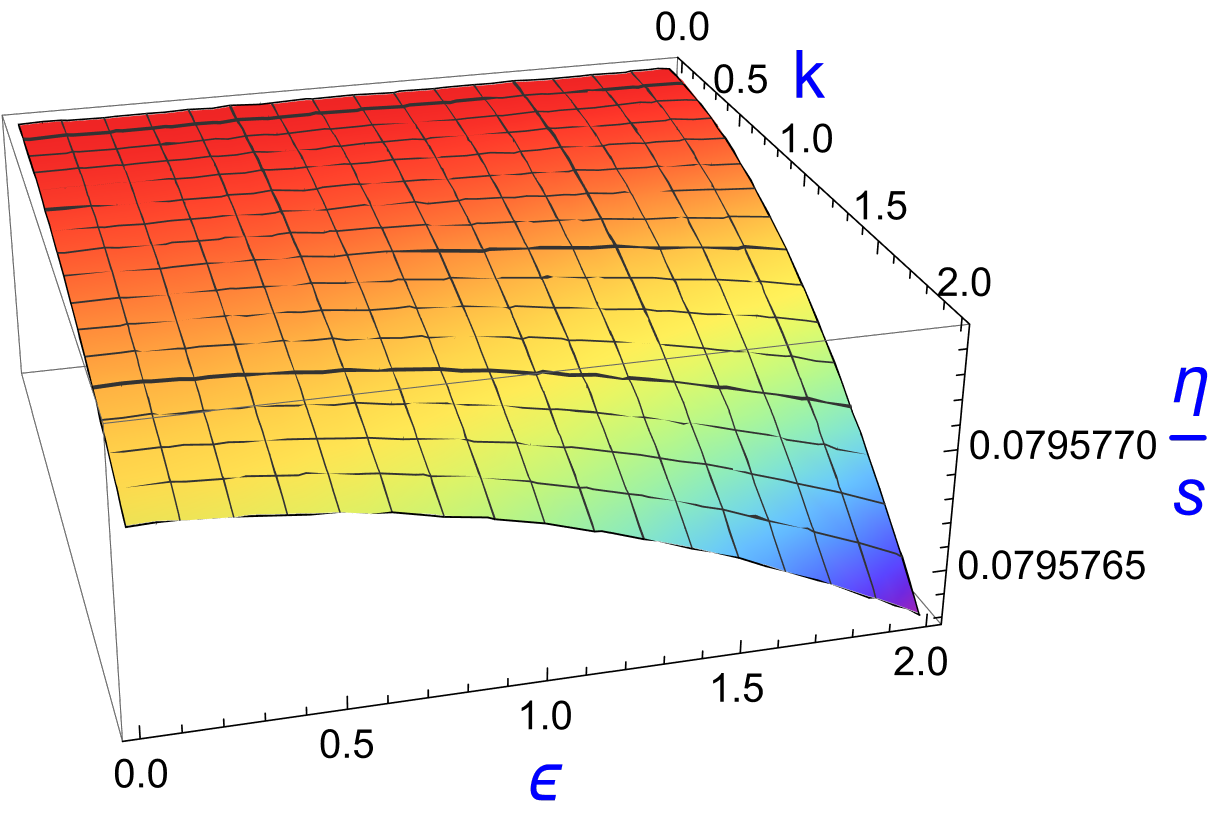}\hspace{0.1cm}
\includegraphics[width=0.49\linewidth,height=4.7cm]{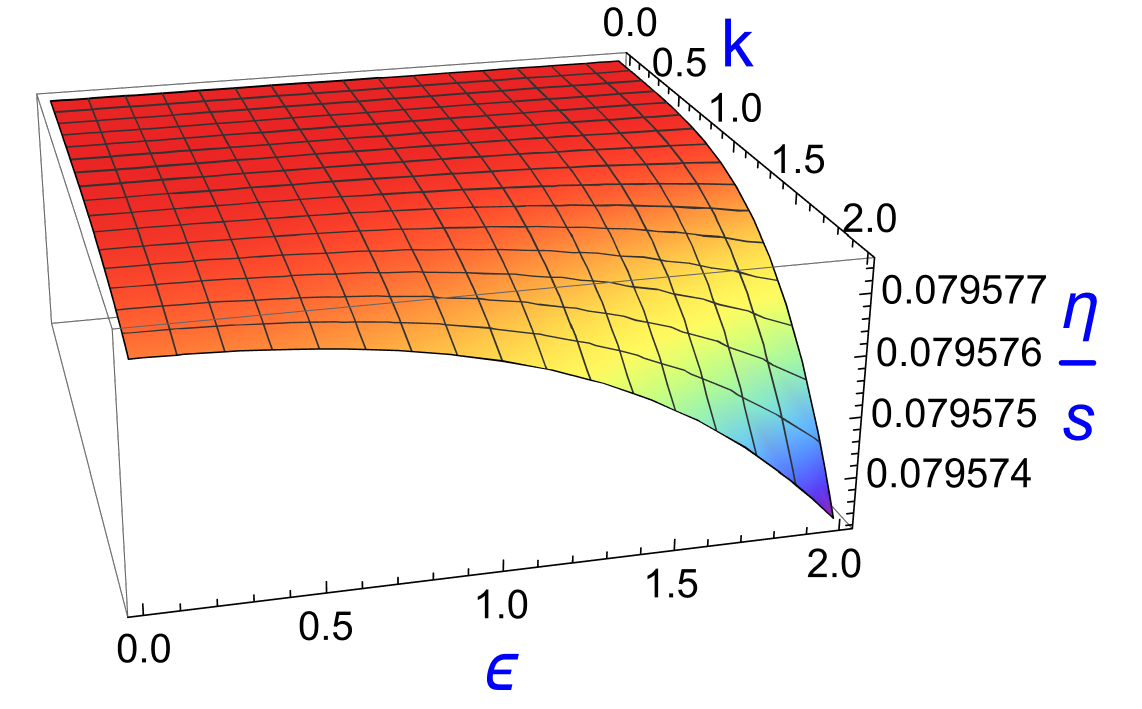}
    \caption{The ratio $\eta/s$ as the function of the $k$ and $\epsilon$, depicted by  (\ref{two-axion eta11}). Here, we have fixed $\lambda^2\,L^2=10^{-5}$ and $r_h/L=2$. $\textbf{Left}$: $n=1$. $\textbf{Right}$: $n=2$. In both cases, the symmetries are explicitly broken.}
    \label{fig2:eta}
\end{figure}
\begin{figure}[htbp]
    \centering  \includegraphics[width=0.49\linewidth,height=4.7cm]{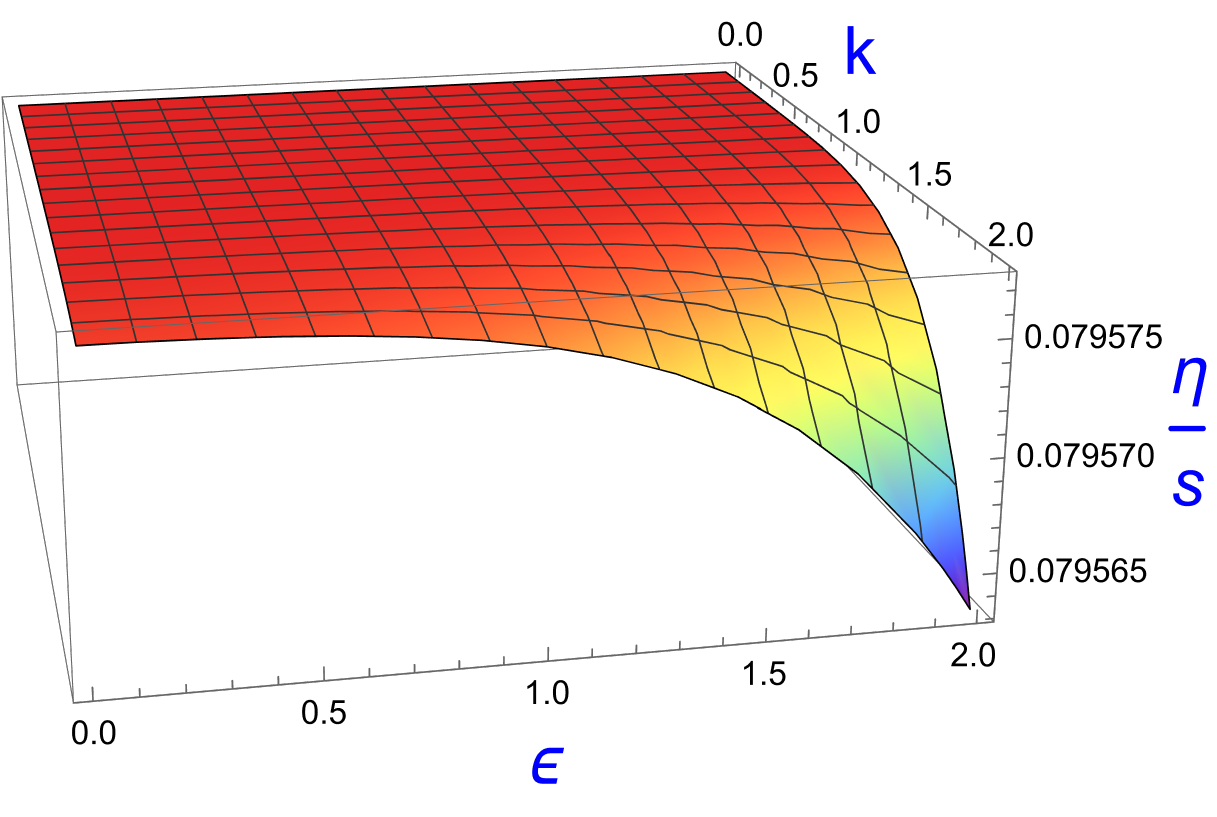} \hspace{0.1cm}  \includegraphics[width=0.49\linewidth,height=4.7cm]{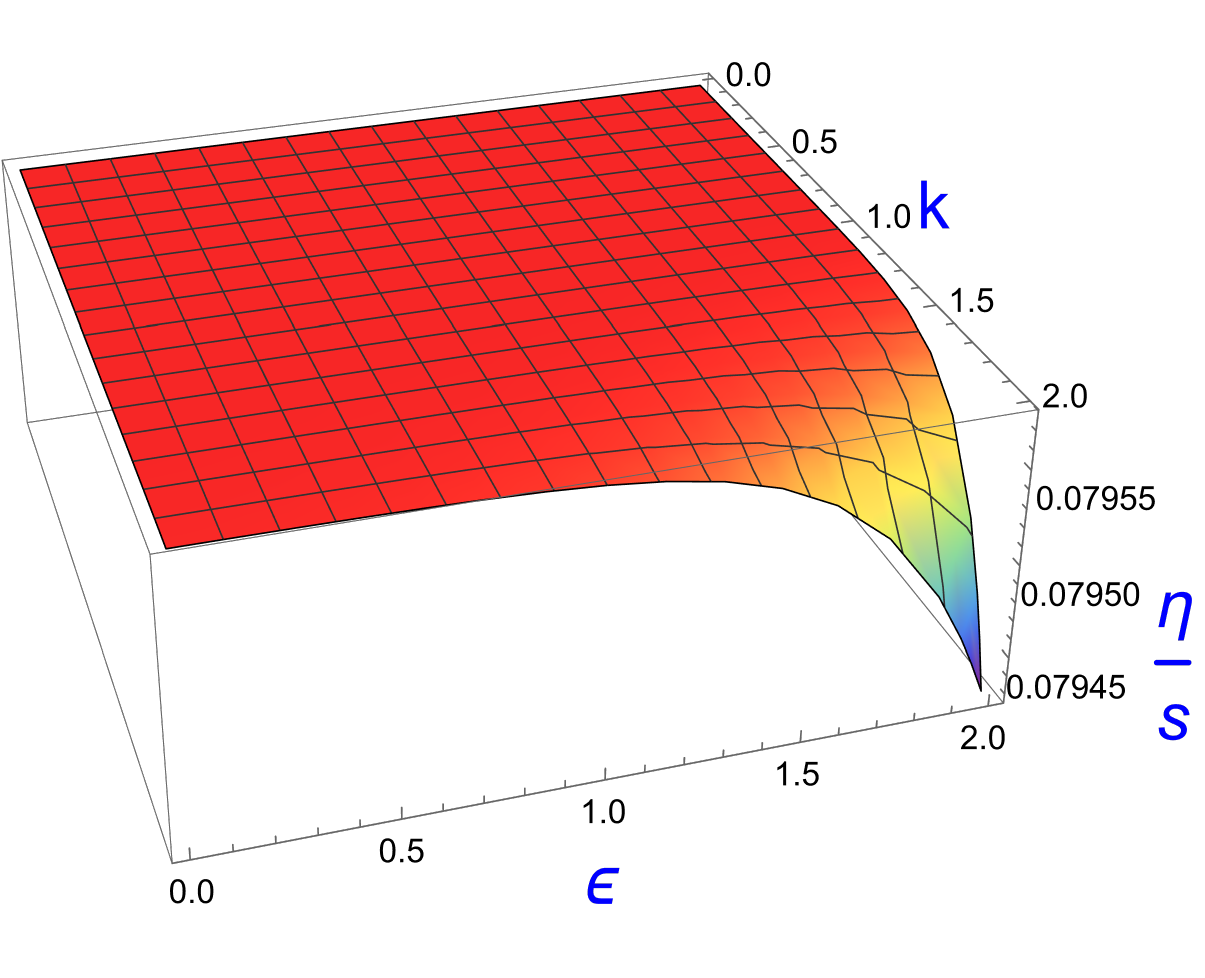}
    \caption{The ratio $\eta/s$ as the function of the $k$ and $\epsilon$, depicted by  (\ref{two-axion eta11}). Here, we have fixed $\lambda^2\,L^2=10^{-5}$ and $r_h/L=2$. $\textbf{Left}$: $n=3$. $\textbf{Right}$: $n=5$. In both cases, the symmetries are spontaneously broken.}
    \label{fig3:eta}
\end{figure}

\section{Discussion and outlook}
In this paper, the shear elasticity and the shear viscosity of an anisotropic holographic axion model with broken translations have been investigated. We find that 
a positive shear modulus emerges when the translational symmetry is broken spontaneously. In particular, the shear modulus is doubly enhanced in the presence of the broken translations and anisotropy.
Furthermore, the $\eta/s$ ratio is always suppressed no matter the translations as well as the rotations are broken explicitly or spontaneously which is in contrast to the holographic p-wave superfluid case. In the p-wave superfluid, the spontaneous breaking of rotational symmetry caused by a vector condensate enhances the value of $\eta/s$, competing against the effects of explicit symmetry breaking. The KSS bound is hence never violated \cite{Baggioli:2023yvc}. From this perspective, we can conclude that the fate of the KSS bound in holographic systems cannot be solely attributed to whether rotational symmetry is broken, or to the specific manner of that breaking (whether spontaneous or explicit). The $\eta/s$ ratio also depends very much on the details of the operator that gives rise to the symmetry breaking.

For case of the spontaneous symmetry breaking with $n>5/2$, the holographic model is dual to a solid on boundary. Then, one may naively expect that the violation of the viscosity bound cannot happen and the shear viscosity should be divergent in a perfect crystal without topological defects, say, the dislocation flow. However, this does not contradict our results. We consider the frequency-dependent viscosity of the crystal given by the Kubo formula \cite{Delacretaz:2017zxd}, $\eta^{\text{crystal}}(\omega)\equiv \text{Im}\,[G^R_{T_{xy}T_{xy}}(\omega,\boldsymbol{p}=0)/\omega]\approx \mu \,\delta(\omega)+\eta$, where the delta function is from the $1/\omega$ divergence of the real part according to the Kramers–Kronig relation
and $\eta$ is the dissipative coefficient addressed in this work. In the DC limit, we have $\eta^{\text{crystal}}\rightarrow \infty$ due to the delta function. Therefore, although the KSS bound is violated, the crystal viscosity receives a divergent contribution from the rigidity of the translational order.

Nevertheless, our study in this work is limited to the perturbative region, i.e., the symmetry breaking is very weak. Therefore, it is not clear whether our conclusion holds or not when the breaking of the symmetries is finite.  In more general case, the anisotropic background solutions cannot be obtained by the analytic way. However, to achieve a complete answer, it is necessary to extend our study by, for instance, following the numeric methods used in \cite{Rebhan:2011vd,Mamo:2012sy}
 to obtain the background solutions for finite values of $\lambda$ and analyze how the viscoelastic property is affected in the strongly anisotropic region. We leave this for future work.

\acknowledgments
We would like to thank Matteo Baggioli, Kang Liu, Yan Liu, Ya-Wen Sun and Ling-Zheng Xia for numerous helpful discussions. This work is supported by the National Natural Science Foundation of China (NSFC) under Grants No.12275038 and No.12375054.

\appendix
\section{Corrections to the background solutions due to the axion fields}\label{Appendix A}

By the perturbative calculation, we can obtain that the general solution of $a_1^{(n)}(r)$  as follows 
\begin{eqnarray}
a_1^{(n)}(r)=\frac{k_y^{2 n}\,r^{2-2 n}}{(2 n-3)\,2^{n+1}}-\frac{c_1}{r}+c_2,
\end{eqnarray}
where $c_1$ and $c_2$ are integral constants which can be fixed as $c_1=\frac{ k_y^{2 n}\,r_h^{3-2 n}}{(2 n-3)\,2^{n+1}}$ and $c_2=0$ so that $a_1^{(n)}(r)$ is regular at the horizon and does not destroy the asymptotic AdS. 

The general solution of $b_1^{(n)}(r)$ is
\begin{eqnarray}
b_1^{(n)}(r)=\frac{1}{3} r^2 \left[\frac{ n\,k_y^{2 n}\,\mathcal{B}\left(\frac{r^3}{r_h^3};1-\frac{2n}{3},0\right)}{(2 n-3)\,2^{n}\,r_h^{2 n}}+\frac{c_3\, \text{log}\left(1-\frac{r_h^3}{r^3}\right)}{r_h^3}+3\,c_4\right],
\end{eqnarray}
with integral constants $c_3$ and $c_4$, the incomplete beta function $\mathcal{B}\left(\frac{r^3}{r_h^3},1-\frac{2n}{3},0\right)$ which is logarithmically divergent at the horizon. Then, we have to choose $c_3 = \frac{ n\, k_y^{2 n} r_h^{3-2 n}}{(2 n-3)\,2^{n}}$ so that the logarithmic divergence from $\mathcal{B}$ can be cancelled. In addition, $\mathcal{B}$ is complex, but when $n\neq\frac{3}{2}\textbf{z}$ and $r>r_h$, its imaginary part is independent of $r$. Then, we can choose $c_4$ properly to remove the imaginary part of $\mathcal{B}$. On the other hand, since $\text{Re}\,\mathcal{B}\left(\infty;1-\frac{2n}{3},0\right)=-\pi\,\text{cot }\left(\frac{2 \pi  n}{3}\right)$, we have to fix 
\begin{equation}
c_4=\frac{n\,\pi\, k_y^{2 n}\,r_h^{-2 n}}{3\,(2 n-3) \,2^n}\,\text{cot }\left(\frac{2 \pi  n}{3}\right)-\frac{ n\,k_y^{2 n}\,}{3\,(2 n-3)\,2^{n}\,r_h^{2 n}}\text{Im}\,\mathcal{B}\left(\frac{r^3}{r_h^3};1-\frac{2n}{3},0\right),
\end{equation}
so that the boundary is still AdS. However, when
$n= \frac{3}{2}\left(\textbf{z}+1\right)$ and $r>r_h$, the real part of $\mathcal{B}$ becomes divergent. Then, we have to choose $c_4$ properly so that $\text{Im}\,\mathcal{B}$ and the divergence of $\text{Re}\,\mathcal{B}$ are both removed. In this case, only the regular part of $\mathcal{B}$ (which we denoted as $\mathcal{B}_0$ in the main text) is left. One can further check that $\text{Re}\,\mathcal{B}_0\left(\infty;1-\frac{2n}{3},0\right)=0$. Then, in all the cases (except for $n=\frac{3}{2}$), we obtain the result (\ref{b1}) in the main text.
 
At the horizon, we obtain that
\begin{equation}\label{btogamma}
b_1^{(n)}(r_h)=
\begin{cases}
\frac{n\, k_y^{2n}\,r_h^{2-2 n}}{3\cdot2^{n}\,(3-2 n)}\mathcal{H}(\frac{2n-3}{3}), & n\neq\frac{3}{2},\, \\
-\frac{\pi ^2 k_y^3}{72 \,\sqrt{2} r_h}, & n= \frac{3}{2}. \\
\end{cases}
\end{equation}

\begin{figure}[htbp]
    \centering    \includegraphics[width=0.6\linewidth,height=7cm]
    {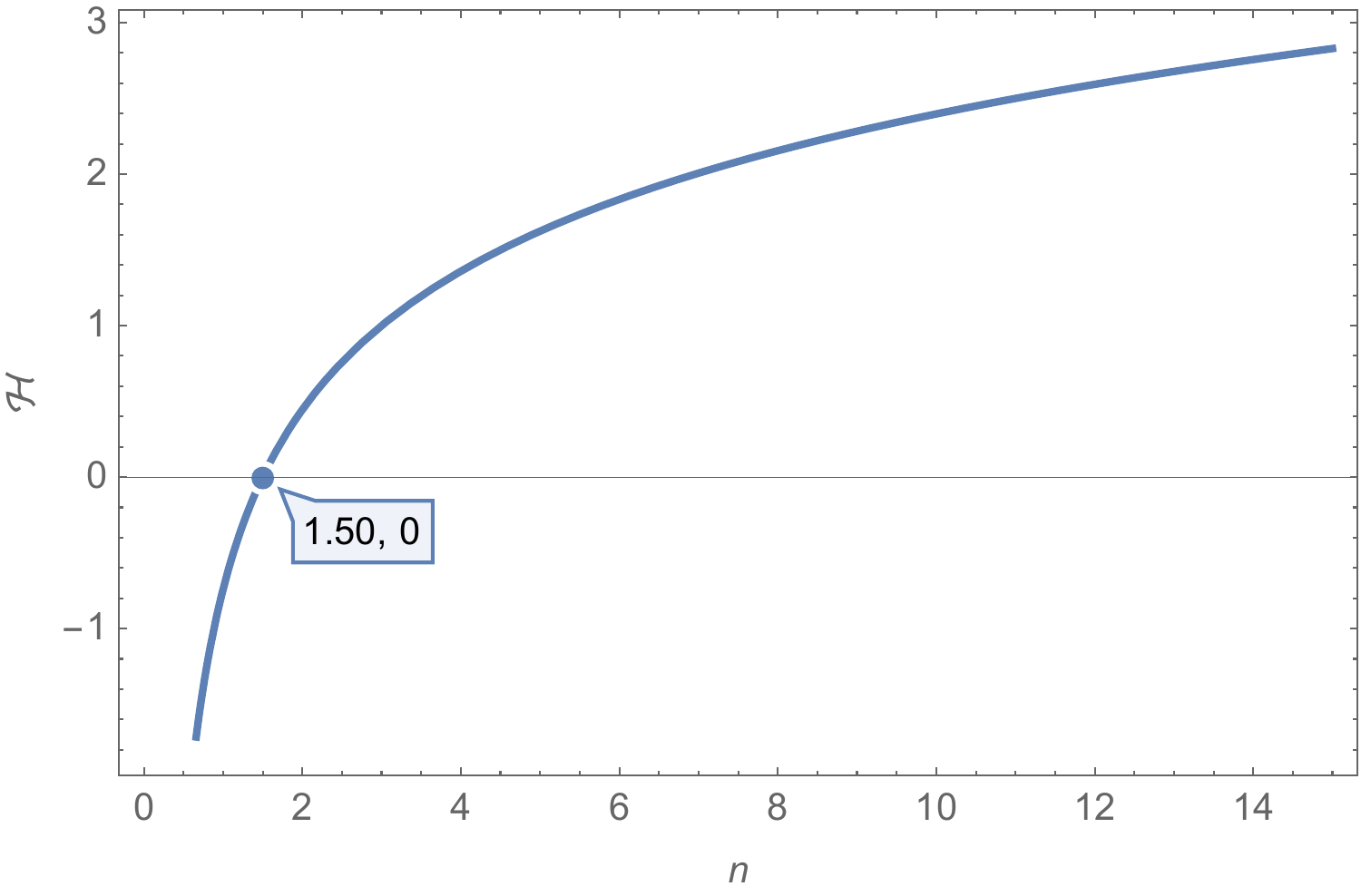}   
    \caption{$\mathcal{H}(\frac{2n-3}{3})$ as the function of $n$. }
    \label{fig:HarmonicNumber}
\end{figure}

In FIG.\ref{fig:HarmonicNumber}, we plot $\mathcal{H}(\frac{2n-3}{3})$ as the function of $n$. When $0<n<\frac{3}{2}$, we have $\mathcal{H}(\frac{2n-3}{3})<0$. In contrast, when $n>\frac{3}{2}$, we find that $\mathcal{H}(\frac{2n-3}{3})>0$. Then, we conclude that $b_1^{(n)}(r_h)$ is always negative in the one-axion case.

Repeating the calculation, in the two-axion case, we have that
\begin{equation}
b_2^{(n)}(r_h)=
\begin{cases}
\frac{n\,r_h^{2-2 n} \left(k_y^2-k_x^2\right)\, \left(k_x^2+k_y^2\right){}^{n-1}}{3\cdot2^{n}\,(3-2 n)}\mathcal{H}(\frac{2n-3}{3}), & n\neq\frac{3}{2},\\
-\frac{\pi ^2 \left(k_y^2-k_x^2\right) \sqrt{k_x^2+k_y^2}}{72 \sqrt{2}\,r_h},& n= \frac{3}{2}. \\
\end{cases}
\end{equation}
Inserting the result above into (\ref{two-axion eta1}) in the main text, we immediately obtain the result (\ref{two-axion eta11}).

\section{Derivation of the shear viscoelasticity for the two-axion case}\label{Appendix B}
We consider the two-axion case and take the fluctuating metric $h^{x}_{y}$ for example and denote it as $h$ for simplicity. Recall that it satisfies
\begin{eqnarray}\label{fluctuation equation}
h''+\left(\frac{A' }{A}+\frac{3 B'}{2 B}-\frac{C' }{2 C}\right)h'+\left(\frac{\omega ^2}{A^2}-\frac{m_g^2}{A}\right) h=0, \,\,\,\,m_g^2=\frac{\lambda^2 k_x^2 V'}{B},
\end{eqnarray}
where $h'$ denotes the derivative of $h$ with respect to $r$.

To compute the Green function of the stress tensor, we look for a special solution to the equation that is normalized to be unit at the AdS boundary. Suppose that the solution near the boundary behaves like
\begin{eqnarray}\label{boundary behavior}
h=(1+\cdots)+h_3(\omega)\,{r^{-3}}(1+\cdots).
\end{eqnarray}
The Green function of the stress tensor can be read off as
\begin{eqnarray}
G^{R}_{T_{xy}T_{xy}}(\omega,\boldsymbol{p}=0)=-3\,h_3(\omega).
\end{eqnarray}
To extract $\eta$ and $\mu$, we need to solve (\ref{fluctuation equation}) up to $\mathcal{O}(\omega)$ by taking the low frequency expansion.  In addition, we require the solution  to satisfy the infalling boundary condition at the horizon. We hence take the following ansatz that
\begin{eqnarray}\label{ansatz}
h(r)= {f(r)}^{-\mathrm{i}\omega/4\pi T}\left[H_0^{(0)}(r)+\lambda^2\,H_0^{(2)}(r)+ \frac{\mathrm{i}\omega}{4\pi T}\Big(H_1^{(0)}(r)+\lambda^2\,H_1^{(2)}(r)\Big)\,+\mathcal{O}(\omega^2)\right].
\end{eqnarray}
where $f(r)\equiv A(r)/r^2$. $H_0^{(0)}(r)$, $H_0^{(2)}(r)$, $H_1^{(0)}(r)$ and $H_1^{(2)}(r)$ should be regular functions in the bulk including the horizon. The upper index of $H(r)$ denotes the order of $\lambda$ and the lower index of $H(r)$ denotes the order of $\omega$.

Firstly, let us calculate the zeroth order (i.e., $m_g^2=0$). Plugging (\ref{ansatz}) into (\ref{fluctuation equation}), we obtain the  following two equations
\begin{align}\label{H00 equation}
0=&{H_0^{(0)}}''+\left(\frac{f' }{f}+\frac{2}{r}+\frac{3 B'}{2 B}-\frac{C' }{2 C}\right){H_0^{(0)}}',
\end{align}
\begin{align}\label{H10 equation}
0=& {H_1^{(0)}}' \left(\frac{f'}{f}+\frac{2}{r}+\frac{3
   B'}{2B}-\frac{C'}{2C}\right)+{H_1^{(0)}}''-{H_0^{(0)}}'\frac{2f'}{f}\\ \nonumber
&-H_0^{(0)}\left[\left(\frac{2}{r}+\frac{3B'}{2B}-\frac{C'}{2C}\right)\frac{f'}{f}+\frac{f''}{f}\right].
\end{align}
The regular solution to (\ref{H00 equation}) that meets the asymptotic behavior (\ref{boundary behavior}) is given by
\begin{eqnarray}\label{H00 boundary}
{H_0^{(0)}}=1.
\end{eqnarray}
With this, $H_1^{(0)}(r)$ can be further determined by (\ref{H10 equation}) and (\ref{boundary behavior}), which is
\begin{eqnarray}\label{H10}
{H_1^{(0)}}(r)=c_6+\int_{{r_{h}}}^{r}\mathrm{d}y\frac{f^{\prime}(y)}{f(y)}+c_5\int_{r_h}^{r} \mathrm{d}y\frac{C(y)^{1/2}}{A(y)B(y)^{3/2}}.
\end{eqnarray}
Combining the regularity of $H_1^{(0)}$ at the horizon and analyzing its asymptotic behavior, one can obtain that $c_5=-\frac{A'(r_h)B(r_h)^{3/2}}{C(r_h)^{1/2}}$ and $c_6=0$ .
Then, near the boundary($r\rightarrow \infty$), we find that
\begin{align}\label{H10 boundary}
&H_1^{(0)}(r)=\frac1{3r^3}\frac{A^{\prime}(r_h)B(r_h)^{3/2}}{C(r_h)^{1/2}}+\cdots,
\end{align}
\begin{align}
&h(r)=1+\frac{\mathrm{i}\omega B(r_h)^{3/2}C(r_h)^{-1/2}}{3r^3}+\cdots.
\end{align}
Using the holographic dictionary, 
the retarded Green function can be read off as
\begin{eqnarray}
G^{R}_{T_{xy}T_{xy}}(\omega,\boldsymbol{p}=0)=-
\mathrm{i}\omega\,B(r_h)^{3/2}C(r_h)^{-1/2}\equiv -\mathrm{i} \omega \eta,
\end{eqnarray}
which is purely imaginary. Since the entropy density $s=4\,\pi\sqrt{BC}|_{r=r_h}$, the viscosity to entropy density ratio should be
\begin{eqnarray}
\frac{\eta}{s}=\frac{1}{4\pi}\frac{B(r_h)}{C(r_h)}.
\end{eqnarray}
If the system has an $SO(2)$ symmetry on the $x-y$ plane, i.e, $B(r)=C(r)$, it reduces to the celebrated KSS bound.

Now, we turn to the order $\mathcal{O}(\lambda^{2})$ by setting ${m_g^2}(r)\equiv \lambda^2\, {M^2(r)}$ with $M^2(r)=\frac{k_x^2 V'}{B}$. Subsequently, we have two equations
\begin{align}\label{H02 equation}
0=&{H_0^{(2)}}''+\left(\frac{f' }{f}+\frac{2}{r}+\frac{3 B'}{2 B}-\frac{C' }{2 C}\right){H_0^{(2)}}'-\frac{M^2}{r^2f} H_0^{(0)},
\end{align}
\begin{align}\label{H12 equation}
0=& {H_1^{(2)}}' \left(\frac{f'}{f}+\frac{2}{r}+\frac{3
   B'}{2B}-\frac{C'}{2C}\right)+{H_1^{(2)}}''-{H_1^{(0)}} \frac{M^2}{r^2f}-{H_0^{(2)}}'\frac{2f'}{f}\\ \nonumber
&-H_0^{(2)}\left[\left(\frac{2}{r}+\frac{3B'}{2B}-\frac{C'}{2C}\right)\frac{f'}{f}+\frac{f''}{f}\right].
\end{align}
We just simply replace the background appearing in order $\mathcal{O}(\lambda^{2})$ with the Schwarzschild metric. We find that the third and last terms in (\ref{H12 equation}) vanish on the Schwarzschild background\footnote{It is found that $H_1^{(0)}$ is zero when we just bring the Schwarzschild metric into (\ref{H10}).}.
In this way, we obtain that
\begin{align}
{H_0^{(2)}}(r)= &c_8+c_7\int^{r}_{r_h}\mathrm{d}x\frac{C(x)^{1/2}}{A(x){B(x)}^{3/2}}\\ \nonumber
&+\int^{r}_{r_h}\mathrm{d}x\left(\frac{C(x)^{1/2}}{A(x){B(x)}^{3/2}}\int_{r_h}^x \mathrm{d}y \frac{{B(y)}^{3/2}{H_0}^{(0)}(y)}{C(y)^{1/2}}{{M^2}(y)}\right),
\end{align}

\begin{align}\label{H12-2}   
{H_1^{(2)}}(r)=&c_{10}+c_{9}\int_{r_{h}}^{r}\mathrm{d}x \frac{C(x)^{1/2}}{A(x)B(x)^{3/2}}\\ \nonumber
&+\int_{r_{h}}^{r}\mathrm{d}x\left( \frac{C(x)^{1/2}}{A(x)B(x)^{3/2}}\int_{r_{h}}^{x}\mathrm{d}y \frac{2y^{2}B(y)^{3/2}f^{\prime}(y)}{C(y)^{1/2}}{H_{0}^{(2)}}'(y)\right).\\ \nonumber
\end{align}

\subsubsection{$n\neq\frac{3}{2}$}
 Using (\ref{H00 boundary}) and ${M^2}(y)=\frac{k_x^2\,V'}{B(y)}=\frac{n\,k_x^2}{y^2}  \left(\frac{k_x^2}{2\,y^2}+\frac{k_y^2}{2\,y^2}\right)^{n-1}$, we obtain
 \begin{eqnarray}
{H_0^{(2)}}(r)&=&\int^{r}_{r_h}\mathrm{d}x\left[\frac{C(x)^{1/2}}{A(x){B(x)}^{3/2}}\int_{r_h}^x \mathrm{d}y\frac{n\,y^{2-2 n}\, k_x^2\left(k_x^2+k_y^2\right)^{n-1}}{\,2^{n-1}}\right]\\ \nonumber
&=&N\int_{r_h}^r\mathrm{d}x \frac{x^{3-2n}}{x^4(1-\frac{r_h^3}{x^3})}-N\,r_h^{3-2n}\int_{r_h}^r\mathrm{d}x \frac{1}{x^4(1-\frac{r_h^3}{x^3})},\,\,\,n\neq\frac{3}{2},
\end{eqnarray}
 where $N\equiv\frac{n\,k_x^2\left({k_x^2+k_y^2}\right)^{n-1}}{2^{n-1}(3-2n)}$. The two conditions of regular function and asymptotic behavior near the boundary require the integral constants $c_7$ and $c_8$ to be zero.
 Then, 
\begin{eqnarray}\label{H02`}
 &&{H_{0}^{(2)}}'(r)=N\frac{r^{3-2n}}{r^4(1-\frac{r_h^3}{r^3})}-N\frac{r_h^{3-2n}}{r^4(1-\frac{r_h^3}{r^3})}.
 \end{eqnarray}
 Near the boundary ($r\rightarrow \infty$),
\begin{eqnarray}\label{H02 boundary}
&&H_{0}^{(2)}(r)=-\frac1{3r^3}\frac{{nk_x^2\left({k_x^2+k_y^2}\right)^{n-1}}}{2^{n-1}(2n-3)}r_h^{3-2n}+\cdots. 
\end{eqnarray}
Combining (\ref{H12-2})  and (\ref{H02`}), we have
\begin{align}\label{H12-3}
{H_1^{(2)}}(r)=&c_{10}+c_{9}\int_{r_{h}}^{r}\mathrm{d}x \frac{C(x)^{1/2}}{A(x)B(x)^{3/2}}+6\,N\,r_h^{3}\int_{r_{h}}^{r}\mathrm{d}x\left[\frac{C(x)^{1/2}}{A(x)B(x)^{3/2}}\Big(F(x)\right.\\\nonumber
&\left.-F(r_h)\Big)\right]-6\,N\,r_h^{6-2n}\int_{r_{h}}^{r}\mathrm{d}x\left[\frac{C(x)^{1/2}}{A(x)B(x)^{3/2}}\Big(G(x)-G(r_h)\Big)\right],
\end{align}
where
\begin{eqnarray}
F(x)-F(r_h)&=&-\frac{1}{3}r_h^{-2n}\Big[\mathcal{B}\left(\frac{y^3}{r_h^3},1-\frac{2n}{3},0\right)\Big]\Bigg|^{y=x}_{y=r_h}, \cr
G(x)-G(r_h)&=&\frac{1}{3\,r_h^3}\Big[\text{log}(\frac{y^3-r_h^3}{y^3})\Big]\Bigg|^{y=x}_{y=r_h},
\end{eqnarray}
with the incomplete beta function $\mathcal{B}\left(\frac{y^3}{r_h^3},1-\frac{2n}{3},0\right)$.
We need to fix the integral constants $c_9=c_{10} =0$ so that $H^{(2)}_1(r)$ ensures the regularity of the solution at the horizon and the asymptotic behavior (\ref{boundary behavior}) near the boundary. Then, near the boundary, we take the expansion of $A(x),B(x),C(x),F(x),G(x)$ and insert them into (\ref{H12-3}). We immediately get
\begin{eqnarray}\label{H12 boundary}
&&H_{1}^{(2)}(r)=-\frac1{3r^3}\frac{{nk_x^2\left({k_x^2+k_y^2}\right)^{n-1}}}{2^{n-2}(2n-3)\,r_h^{2n-3}}\,\mathcal{H}(\frac{2n-3}{3}) +\cdots.
\end{eqnarray}
In the end, for $V(X) =X^n$ model, substituting (\ref{H00 boundary}), (\ref{H10 boundary}), (\ref{H02 boundary}) and (\ref{H12 boundary}) into (\ref{ansatz}), we derive that
\begin{eqnarray}
&& \mu=\lambda^2\frac{n\,k_x^2\,\left({k_x^2+k_y^2}\right)^{n-1}}{2^{n-1}\, (2n-3)}r_h^{3-2n}+\mathcal{O}(\lambda^{4}), \quad n>\frac{3}{2},\\
&&\frac{\eta}{s}=\frac{1}{4\pi}\frac{B(r_{h})}{C(r_{h})}-\lambda^2\frac{n\,  k_x^2\,\left({k_x^2+k_y^2}\right)^{n-1}}{3\cdot2^{n}\,(2n-3)\, \pi\,r_h^{2n}}\,\mathcal{H}(\frac{2n-3}{3})+\mathcal{O}(\lambda^{4})\label{eq:etaOs-n no=3/2}.
\end{eqnarray}

\subsubsection{$n=\frac{3}{2}$}
In this case, ${M^2}(y)=\frac{k_x^2\,V'}{B(y)}=\frac{3\,k_x^2}{2\,y^2}  \left(\frac{k_x^2}{2\,y^2}+\frac{k_y^2}{2\,y^2}\right)^{1/2}$, we have 
\begin{eqnarray}\label{n=3/2,H02}
{H_0^{(2)}}(r)=\frac{3 \,k_x^2 \sqrt{k_x^2+k_y^2}}{2 \sqrt{2}}\int_{r_h}^r\mathrm{d}x \frac{\text{log}(\frac{x}{r_h})}{x^4(1-\frac{r_h^3}{x^3})},
\end{eqnarray}
where $c_7$ and $c_8$ are also zero. Then, 
\begin{eqnarray}\label{n=3/2,H02`}
 &&{H_{0}^{(2)}}'(r)=\frac{3 \,k_x^2 \sqrt{k_x^2+k_y^2}}{2 \sqrt{2}}\frac{\text{log}(\frac{r}{r_h})}{r^4(1-\frac{r_h^3}{r^3})}.
 \end{eqnarray}
Near the boundary ($r\rightarrow \infty$),
\begin{eqnarray}\label{n=3/2,H02 boundary}
H_{0}^{(2)}(r)=-\frac{1}{r^3}\frac{k_x^2 \sqrt{k_x^2+k_y^2}}{6 \sqrt{2}}+\frac{1}{r^3}\,\text{log}(\frac{r_h}{r})\frac{k_x^2 \sqrt{k_x^2+k_y^2}}{2 \sqrt{2}}+\cdots,
\end{eqnarray}
where  $\text{log}(\frac{r_h}{r})$ is divergent at the boundary.

Combining (\ref{H12-2})  and (\ref{n=3/2,H02`}), near the boundary, the regular function $H_{1}^{(2)}(r)$ is obtained by repeating the calculation
\begin{eqnarray}\label{n=3/2,H12 boundary}
&&H_{1}^{(2)}(r)=-\frac{1}{r^3}\frac{\pi^2\,k_x^2 \sqrt{k_x^2+k_y^2}}{18 \sqrt{2}} +\cdots,
\end{eqnarray}
where we have chosen $c_{9}=c_{10}=0$ so that $H_{1}^{(2)}(r)$ ensures the regularity at the horizon and the asymptotic behavior (\ref{boundary behavior}) near the boundary.
Finally, for $V(X) =X^\frac{3}{2}$ model, we find that
\begin{eqnarray}\label{eq:etaOs-n=3/2}
&&\frac{\eta}{s}=\frac{1}{4\pi}\frac{B(r_{h})}{C(r_{h})}-\lambda^2 \frac{\pi\,k_x^2 \sqrt{k_x^2+k_y^2}}{72\,\sqrt{2}\,r_h^3}+\mathcal{O}(\lambda^{4}).
\end{eqnarray}

\section{Shear viscosity calculated by the eom of $h^{y}_{x}$}\label{Appendix C}

In the section of one-axion case in the main text, we mentioned that the result of the shear viscosity does not depend on whether solving the eom of $h^x_y$  or the one of $h^y_x$. Here, we will show that solving the equation of the massive $h^y_x$ also gives the result (\ref{one axion eta}). Analogous to the calculations in Appendix \ref{Appendix B}, we have the following formula for the massive field
\begin{equation}\label{etayx}
\frac{\eta}{s}=\frac{1}{4\pi}\frac{C(r_{h})}{B(r_{h})}-
\begin{cases}
\lambda^2\frac{n\,  k^{2n}\,\mathcal{H}(\frac{2n-3}{3})}{3\cdot2^{n}\,(2n-3)\, \pi\,r_h^{2n}}+\mathcal{O}(\lambda^{4}), & n\neq\frac{3}{2}, \\
\lambda^2 \frac{\pi\,k^3}{72\,\sqrt{2}\,r_h^3}+\mathcal{O}(\lambda^{4}),& n= \frac{3}{2},\\
\end{cases}
\end{equation}
which is obtained by setting $k_x=0$ and $k_y=k\neq 0$. This seems very different from the result we achieved in the main text, but expanding 
\begin{equation}\label{CTOB}
\frac{C(r_{h})}{B(r_{h})}=1-\lambda^2\frac{2\,b_1^{(n)}(r_h)}{ r_h^2}+\mathcal{O}(\lambda^4)
\end{equation}
and combining it with (\ref{btogamma}) directly give us
\begin{equation}\label{etayx}
\frac{\eta}{s}=\frac{1}{4\pi}+\lambda^2\,\frac{b_1^{(n)}(r_h)}{2\,\pi\,{r_h}^2}+\mathcal{O}(\lambda^4),
\end{equation}
which is exactly the result (\ref{one axion eta}).

\bibliographystyle{utphys}
\bibliography{reference}

\end{document}